# Improving traffic flow at a 2-to-1 lane reduction with wirelessly connected, adaptive cruise control vehicles

L. C. Davis

10244 Normandy Dr.

Plymouth, Michigan 48170

Wirelessly connected vehicles that exchange information about traffic conditions can reduce delays caused by congestion. At a 2-to-1 lane reduction, the improvement in flow past a bottleneck due to traffic with a random mixture of 40% connected vehicles is found to be 52%. Control is based on connected-vehicle-reported velocities near the bottleneck. In response to indications of congestion the connected vehicles, which are also adaptive cruise control vehicles, reduce their speed in slowdown regions. Early lane changes of manually driven vehicles from the terminated lane to the continuous lane are induced by the slowing connected vehicles. Self-organized congestion at the bottleneck is thus delayed or eliminated, depending upon the incoming flow magnitude. For the large majority of vehicles, travel times past the bottleneck are substantially reduced. Control is responsible for delaying the onset of congestion as the incoming flow increases. Adaptive cruise control increases the flow out of the congested state at the bottleneck. The nature of the congested state, when it occurs, appears to be similar under a variety of conditions. Typically 80-100 vehicles are approximately equally distributed between the lanes in the 500-m region prior to the end of the terminated lane. Without the adaptive cruise control capability, connected vehicles can delay the onset of congestion but do not increase the asymptotic flow past the bottleneck. Calculations are done using the Kerner-Klenov three-phase theory, stochastic discrete-time model for manual vehicles. The dynamics of the connected vehicles is given by a conventional adaptive cruise control algorithm plus commanded deceleration. Because time in the model for manual vehicles is discrete (one-second intervals), it is assumed that the acceleration of any vehicle immediately in front of a connected vehicle is constant during the time interval, thereby preserving the computational simplicity and speed of a discrete-time model.

1. Introduction

The dynamics of vehicular traffic flow has been investigated extensively [1-3], in part due to its relationship to self-organization and non-equilibrium phase transitions. From a societal perspective, the economic cost of congestion motivates research to improve traffic efficiency. With the advent of autonomous vehicles that eliminate reaction times of human drivers and other behavioral characteristics, prospects for increased flow and reduced delays appear good. Furthermore, the widespread adoption of wireless communication between vehicles (and infrastructure) potentially offers substantial improvements. However, ways to use these new capabilities, the so-called connected vehicle systems, remain incompletely explored.

Representative papers relevant to connected vehicles include the following. Schönhof *et al*. [4, 5] studied information flow and detection of traffic jam fronts with wireless communication. Thiemann *et al*. [6] modeled the longitudinal hopping of information for various vehicle trajectories. Kerner *et al*. [7] described potential enhancements in traffic efficiency using wireless communication. Orosz and collaborators [8, 9] analyzed the effects of communication delays on the stability of connected vehicle systems. Technical advances have expanded the capabilities of vehicles so that wirelessly connected autonomous cars could be commonplace in the near future [10-14].

The purpose of this paper is to demonstrate how wireless communication and control without any sensors embedded in the roadway can improve flow at a bottleneck caused by a 2-to-1 lane reduction, which is discussed in Sec. 2 and related simulations are presented in Sec.3. The nature of the congested state near the bottleneck is examined in Sec. 4; while Sec. 5 pertains to connected, but manually driven vehicles. My conclusions are in Sec. 6.

2. Model for bottleneck at a 2-to-1 lane reduction

A prototypical traffic bottleneck is the 2-to-1 lane reduction of a highway [15-17]. Vehicles in the terminated lane must change to the continuous lane prior to reaching the end of the lane. When traffic flow is light, no congestion happens. However, when the incoming flow increases, congestion can be self-organized even when the total flow is substantially less than the capacity of a single lane. Congestion occurs because drivers do not change lanes in a system-optimal manner. Yamauchi *et al.* [18, 19] have described this situation as having a prisoner's dilemma game structure.

I consider the use of wirelessly connected vehicles (WCV) for improving flow at a 2-to-1 lane reduction of a highway. The idea is to mix in randomly a fraction $f$ of WCV to mitigate self-organized congestion. I assume that the WCV are also adaptive cruise control vehicles [20, 21].

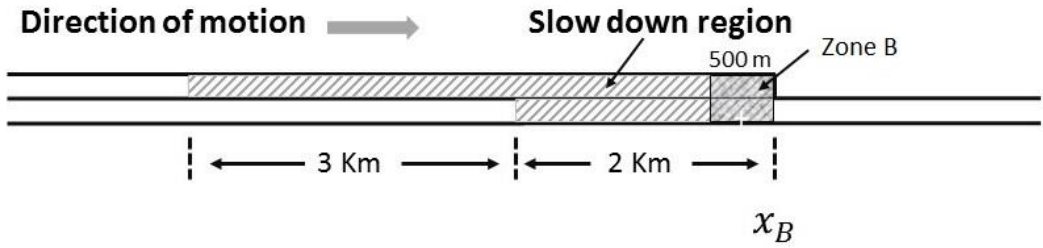

**Fig. 1.** Diagram of a 2-to-1 lane reduction of a highway. The end of the terminated lane is at $x_B$. Slowdown sections for each lane are shown. The region from $x_B$ -500 m to $x_B$ is denoted Zone B.

The lane configuration is depicted in Fig. 1. One lane ends at $x_B$ while the other continues. Prior to the bottleneck is a section of each lane denoted as the "slowdown region" [22]. In these sections, the WCV are instructed to decelerate if preceding vehicles detect congestion building up and communicate this information to others. The current value of the minimum speed of the WCV (self-measured) is reported and a time-dependent function is calculated (possibly by the infrastructure):

$$v_{slow}(t) = max\left\{v_{slow}^{min}, min\{v_i(t)\}\right\}, \tag{1}$$

where $i$ is the index of any WCV. When $x_B - L < x_i \leq x_B$ and $v_i > v_{slow}$ the vehicle $i$ moves according to

$$\frac{dv_i}{dt} = min\{a_{decel}, a_i^d\}, \tag{2a}$$

$$a_{decel} = -0.1 \text{ m/s}^2, \tag{2b}$$

$$a_i^d = \alpha \left[\frac{x_{lead} - x_i - D}{h_d} - v_i\right] + k_d(v_{lead} - v_i). \tag{2c}$$

Vehicle velocities are bounded by the speed limit $v_{lim}$ = 32 m/s. Here $v_{slow}^{min}$ =20 m/s, $L$ = 5000 m (terminated lane) or 2000 m (continuous lane). No exhaustive search to optimize these parameters was attempted because the values used seemed to work well. The other parameters are $D = 7.5$ m, $\alpha = 2$ s$^{-1}$, $h_d = 1$ s, $k_d = 1$ s$^{-1}$. Manual vehicles are assumed to move according to the Kerner-Klenov (KK) stochastic discrete time model [23-25] (See Appendix A.); and the rules for lane changes of all vehicles are given by this model with the exception that the WCV do not change from the continuous to the terminated lane. Because the KK model uses discrete times (one-second update time), it is assumed that for any of the WCV the vehicle immediately in front (labelled "$lead$") moves during a time interval $t_n$ to $t_{n+1}$ with constant acceleration that is given by $\frac{v_{lead}(t_{n+1}) - v_{lead}(t_n)}{t_{n+1} - t_n}$.

### 3. Simulations of flow

For simulations reported in this paper, the incoming flow is shown in Fig. 2. After the maximum flow is reached, the flow remains constant at this value. Each lane has the same incoming flow, the maximum of which is slightly less than half of the capacity of a single lane, namely $\frac{v_{lim}}{h_d v_{lim} + D}$ = 0.810 vehicles/s. Thus, the continuous lane has the capacity to carry the entire incoming flow.

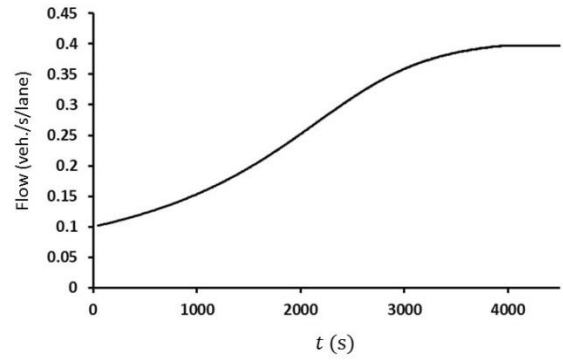

**Fig. 2.** Incoming flow of vehicles as a function of time, which is the time for vehicles to reach $x_B$ from their initial positions traveling at the speed limit of 32 m/s. Each lane of Fig. 1 has the same incoming flow. After the maximum flow is attained at $t \approx 4000$ s, flow remains at this value (0.397 veh./s/lane) thereafter.

In Fig. 3, the positions of lane changes from the terminated lane to the continuous lane are shown as a function of time. The red circles are for all manual flow ($f = 0$) and the blue circles are for flow when 30% of the vehicles are WCV ($f = 0.3$). Many of the lane changes for the all manual flow occur near the bottleneck, whereas the effect of the WCV is to induce earlier lane changes which are beneficial.

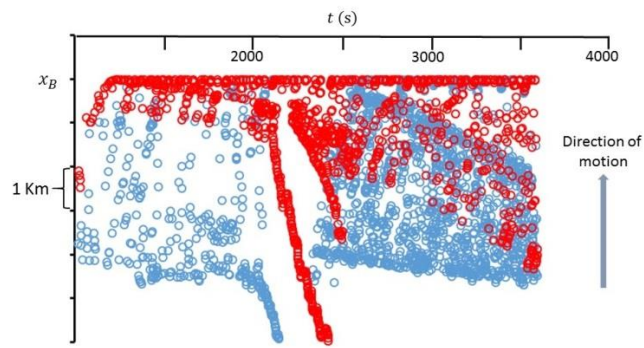

**Fig. 3.** Location of lane changes (from terminated to continuous lane) as a function of time. Red circles are for flow with all manually driven vehicles and blue circles are for mixed traffic with 30% wirelessly connected vehicles.

The number of vehicles $N$ (total of both lanes) in the 500-m region just prior to the bottleneck at $x_B$, called Zone B, is shown in Fig. 4 for all manually driven vehicles and flow with 30% WCV. Ten runs with different random seeds are calculated for each. As the incoming flow increases, congestion sets in at ~1500 s for all manual flow. Flow with 30% WCV remains uncongested until at least 2000 s and in many runs close to 2800 s.

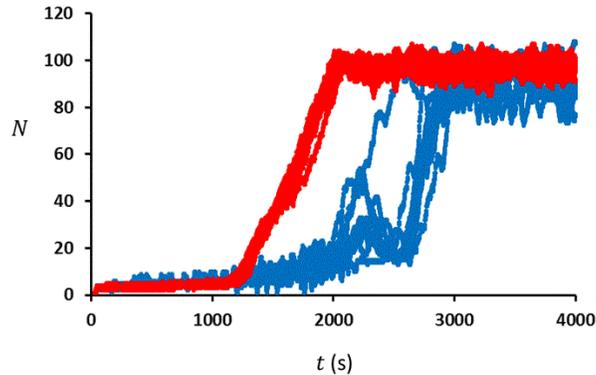

**Fig. 4.** The number of vehicles $N$ in Zone B for flow of all manually driven vehicles (red) and for flow consisting of 30% wirelessly connected vehicles (blue). Ten runs with different random seeds are shown for each.

As expected, the flow past the bottleneck is larger for larger $f$. The average flow during 3000 s to 4000 s is plotted against $f$ in Fig. 5. With $f = 0.4$, the flow is 0.476 s$^{-1}$, which is 59% of capacity, compared to only 0.313 s$^{-1}$ (39% of capacity) for $f = 0$. This represents an improvement of 52% although it falls short of reaching full capacity flow. Simulations (not shown) for nearly all WCV flow suggest that full capacity might be attainable. A more sophisticated algorithm (such as zipper or cooperative merging [26-28]) probably would be employed in this situation. As discussed later, primarily the adaptive cruise control capability of the WCV is responsible for the higher flow after congestion sets in.

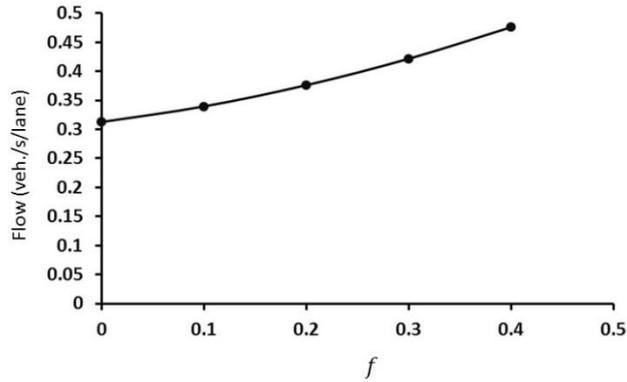

**Fig. 5.** Flow past the bottleneck during 3000 s $< t <$ 4000 s as a function of the fraction $f$ of wirelessly connected vehicles. Average of ten runs shown for each fraction.

## 4. The congested state

In this section, I examine the congested state near the bottleneck by calculating the average velocity $v_{ave}$ of vehicles, as well as the number $N$, in Zone B as shown in Fig. 6. The function $v_{slow}(t)$ is also depicted. The penetration of WCV is 30% ($f = 0.3$). During the first 2000 s, the average velocity $v_{ave}$ closely follows $v_{slow}(t)$ and the number of vehicles $N$ slowly increases to about 20. Thereafter, $N$ increases abruptly and $v_{ave}$ drops below 10 m/s. Near $t = 2500$ s, however, congestion temporarily clears somewhat ($N$ decreases and $v_{ave}$ rises) before the transition to the final congested state where $N \sim$ 80–100 vehicles. I call this initial, temporary congestion the "pre-peak". The gray dashed line is for WCV without deceleration control as given by Eq. 2b. Deceleration control delays the onset of congestion. The adaptive cruise control aspect of the WCV is responsible for the increased flow out of the congested state (Fig. 5). The asymptotic flow is a few percent larger for WCV with deceleration control than without (0.418 compared to 0.408 veh./s).

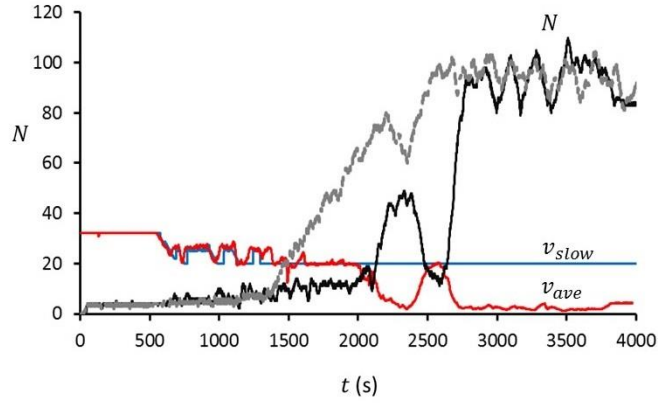

**Fig. 6.** For 30% wirelessly connected vehicles ($f = 0.3$) in both lanes, the function $v_{slow}(t)$ (blue line), the average velocity $v_{ave}$ (red line), and the number of vehicles $N$ (black line) in Zone B. The gray dashed line is $N$ when there is no deceleration control of the connected vehicles. Total incoming flow at large time is $2\,(v_{lim}/\Delta x_{min})$ where $\Delta x_{min} = 80.6$ m, which is the value of the separation (headway plus vehicle length) of vehicles at $t = 4000$ s in Fig. 2.

Fig. 7 depicts $N$ for different final incoming flow. The flow remains constant after it reaches $v_{lim}/\Delta x_{min}$ where $\Delta x_{min}$ is the headway plus $D$ (front bumper to front bumper distance). See Fig. 7b. The flow in Fig. 2 corresponds to $\Delta x_{min} = 80.6$ m. At $\Delta x_{min} = 160$ m, no transition to the congested state is observed (however, see below). For $\Delta x_{min} = 140$ and 120 m, a transition to the congested state occurs and at $\Delta x_{min} = 100$ a pre-peak (as in Fig. 6) is observed before the transition.

The stochastic nature of the pre-peak is shown in Fig. 8. Ten runs are depicted (with different random seeds) with widely different transitions to the congested state. In some runs, no large pre-peak occurs prior to the transition to the congested state, which takes place as early as 2500 s and as late as 3100 s. In contrast, when $\Delta x_{min} = 160$ m, only one of the ten runs exhibits a transition to the congested state (Fig. 9). In this case, the random sequence of WCV contained several large gaps. If all the vehicles are manual, the transition occurs at roughly the same time (~200-s spread) in each run. See Fig. 10.

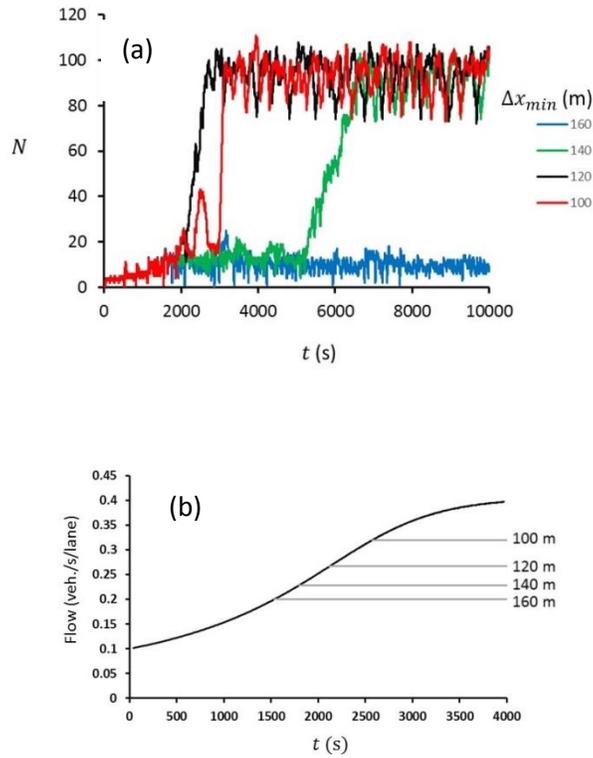

**Fig. 7.** (a) With 30% wirelessly connected vehicles ($f = 0.3$) in both lanes, the number of vehicles $N$ in Zone B for $\Delta x_{min}$ = 100 m (red), 120 m (black), 140 m (green), and 160 m (blue). (b) Incoming flow in each lane which is constant after listed $\Delta x_{min}$ is reached.

The effect of different random sequences is further illustrated in Fig. 11 for $\Delta x_{min}$ = 140 m. Ten runs with random sequences of 30% WCV and ten runs with a regular sequence (where every third vehicle is a connected vehicle) are shown. The regular sequence produces no transition to the congested state, whereas runs with the random sequences show transitions at widely different times.

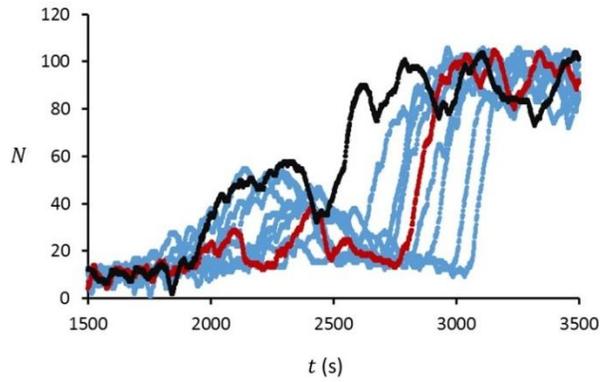

**Fig. 8.** The number of vehicles $N$ in Zone B for $\Delta x_{min} = 100$ m for 10 different runs with 30% wirelessly connected vehicles ($f = 0.3$) in both lanes. For illustration, two individual curves (black and red) are shown demonstrating different characteristics prior to the transition to maximum congestion ($N \sim 80$ to 100 vehicles) due to the stochastic properties of the Kerner-Klenov model (dynamics and lane changing) [23-25].

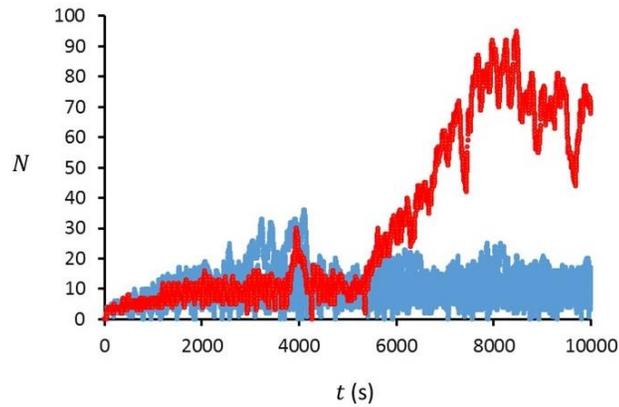

**Fig. 9.** The number of vehicles $N$ in Zone B for $\Delta x_{min} = 160$ m for 10 different runs with 30% wirelessly connected vehicles in both lanes. Only one in ten runs (red curve) shows a transition to the congested state. This occurrence is attributed primarily to a different random sequence of wirelessly connected vehicles.

The congested state appears to be roughly the same regardless of how it is obtained. The number of vehicles $N$ in the region just before the bottleneck (Zone B) in every run fluctuates in the approximate range of 80 to 100 vehicles. The makeup of the congested state is illustrated in Fig. 12. About half of the vehicles are in each lane, although the majority of the WCV are in the continuous lane as might be expected. There are usually slightly more vehicles in the terminated lane compared to the continuous lane.

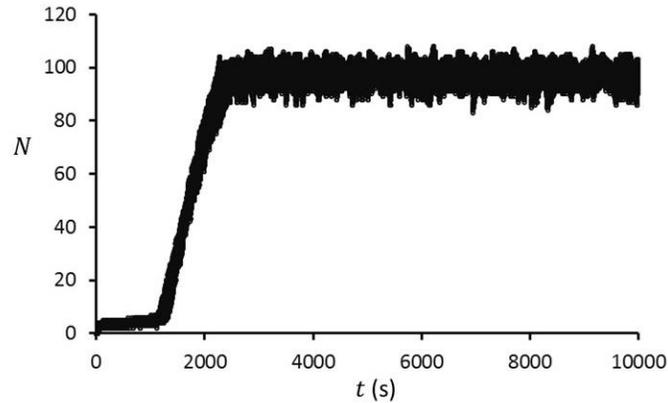

**Fig. 10.** The number of vehicles $N$ in Zone B for $\Delta x_{min} = 160$ m for 10 different runs with only manual vehicles.

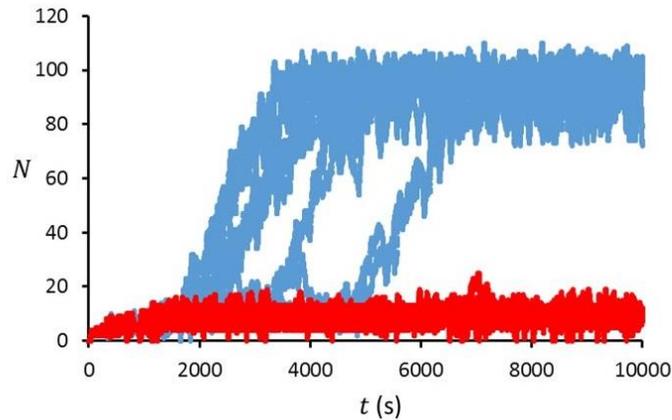

**Fig. 11.** The number of vehicles $N$ in Zone B for $\Delta x_{min} = 140$ m for 10 different runs with random sequences of 30% wirelessly connected vehicles (blue curves) and for 10 runs with a regular sequence (every third vehicle) of wirelessly connected vehicles (red curves).

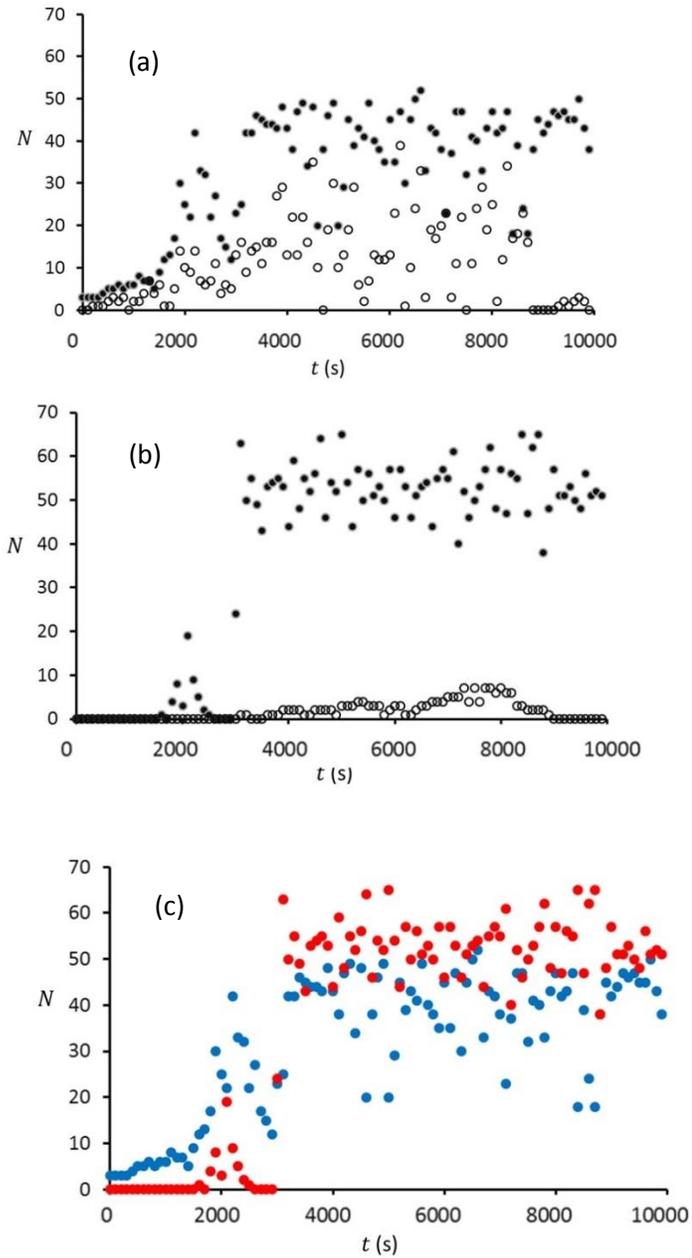

**Fig. 12.** The number of vehicles $N$ in Zone B for $\Delta x_{min} = 100$ m for 30% wirelessly connected vehicles. For clarity, only one run is shown. The full circles are for all vehicles and the open circles are only for wirelessly connected vehicles in the continuous lane (a) and the terminated lane (b). In panel (c), blue (red) dots denote all vehicles in the continuous (terminated) lane.

## 5. Effects of information, deceleration control and adaptive cruise control

The improvement in traffic flow past the bottleneck discussed here is due to a combination of: (1) reducing speed to limit incoming flow, similar to variable speed limit control [29-33]; (2) inducing lane changes to the continuous lane, similar to optimal lane selection [34]; and (3) the adaptive cruise control dynamics of WCV, which is known to improve string stability [20].

In this section, for comparison, I begin by eliminating the adaptive cruise control aspect and consider a scenario where all vehicles are manually driven but receive information and deceleration control [Eq. (2b)] instructions (assuming complete compliance). Also only lane changes to the continuous lane are permitted. The transition to the congested state is displayed in Fig. 13a. The number of vehicles $N$ near the bottleneck rises sharply and their average speed $v_{ave}$ drops to near zero at about $t = 2700$ s. The speed limit in the slowdown regions is $v_{slow} = 20$ m/s throughout. As shown in Fig. 13b, the transition takes place at about $t = 1700$ s without any controls, which is 1000 s sooner than with controls and information.

In Fig. 13c the total number $N_{total}$ of vehicles passing the bottleneck as a function of time is shown for the conditions of Fig. 13a and for all manual vehicle flow with no information or control. The flow past $x_B$ for $t > 4000$ s is nearly the same for each case, namely 0.32 veh./s. The primary effect of information and control appears to be to delay the onset of the transition but not to alter the eventual flow. A similar effect occurs when information and deceleration control are eliminated (but adaptive cruise control is kept) for flow with 30% WCV. In Figs. 6 and 13, $N$ increases much sooner when there is no deceleration control.

It is interesting to compare the distribution of vehicles between the lanes near the bottleneck for all manual flow with no controls (Fig. 14) to the congested state depicted in Fig. 12c, which includes WCV. In each case, the terminated lane has ~60 vehicles. The continuous lane has a broader distribution of the number of vehicles in Fig. 12c compared to Fig. 14, but nearly the same average value. For all the conditions investigated in this work, the congested state reached is remarkably similar, especially the number of vehicles $N$ in Zone B.

Finally, I compute the time delay $T$ for each vehicle in flow of all manual vehicles and in flow with 30% WCV (See Fig. 15a.). The time delay for each vehicle is determined from the distance $d$ travelled in $10^4$ s compared to the distance $d_0$ if it could be travelled at the speed

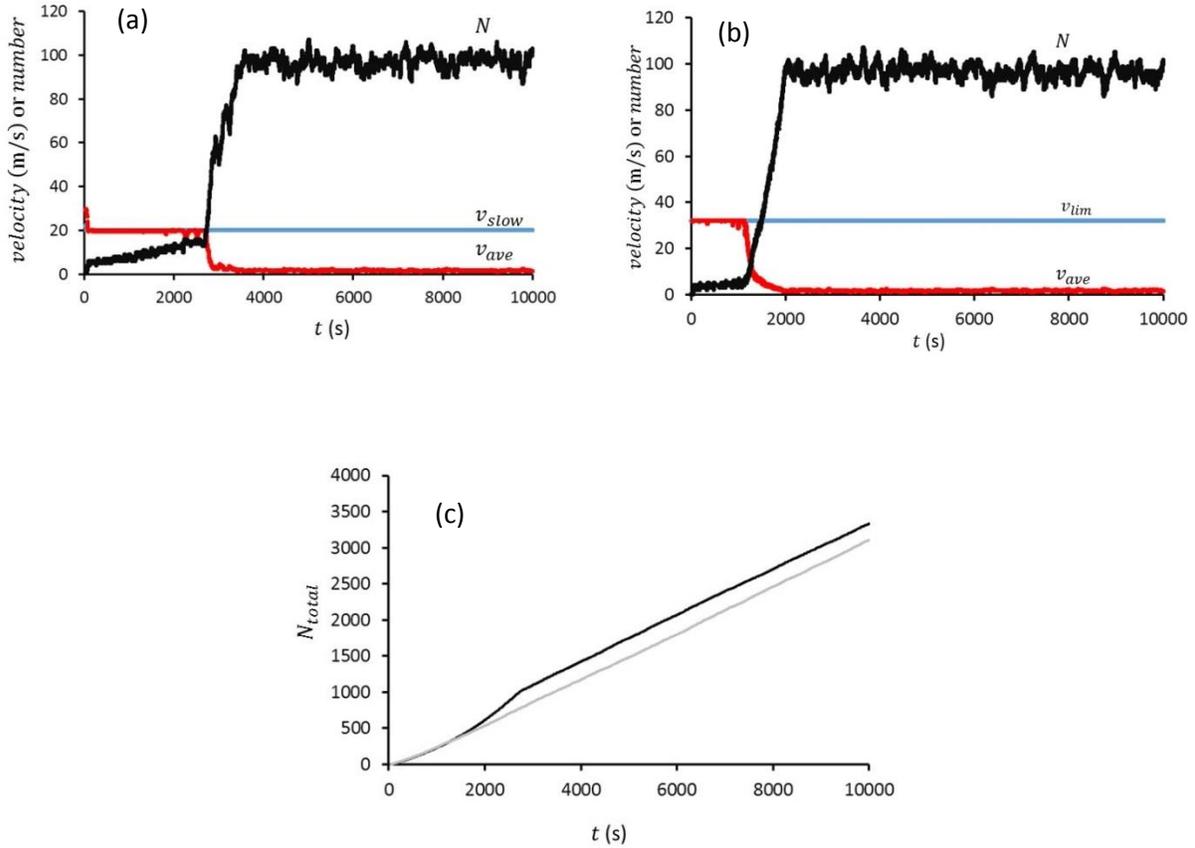

**Fig. 13**. (a) The function $v_{slow}(t)$ (blue line), the average velocity $v_{ave}$ (red line), and the number of vehicles $N$ (black line) in Zone B under conditions where all vehicles are wirelessly connected and obey deceleration instructions but with their dynamics given by the Kerner-Klenov model [23-25], not an adaptive cruise control model. No lane changes from the continuous lane into the terminated lane are allowed. (b) The number of vehicles $N$ in Zone B (black curve) and their average velocity $v_{ave}$ (red curve) with the speed limit (blue line) for all manual vehicle flow with no information or control. (c) The total number $N_{total}$ of vehicles passing $x_B$ vs. time. The black curve corresponds to (a) whereas the gray line is for flow of all manual vehicles with no wireless connections or restrictions on lane changes. The flow (slope of curve) for each is about 0.32 vehicles/s and the total incoming flow is 0.79 vehicles/s for $t >$ 4000 s.

limit, assuming that the vehicle has passed $x_B$: $T = \frac{d_0 - d}{v_{lim}}$. For the majority of vehicles, the time delay for flow with all manual vehicles is larger. The difference $\Delta T$ for each vehicle is plotted in

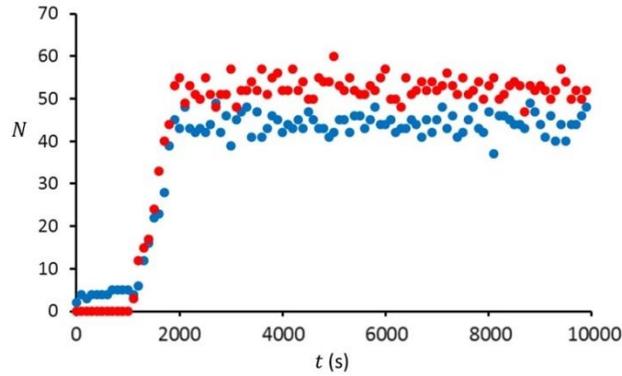

**Fig. 14.** The number of vehicles in Zone B under conditions of Fig. 13b. The blue dots are for vehicles in the continuous lane and the red dots are for the terminated lane.

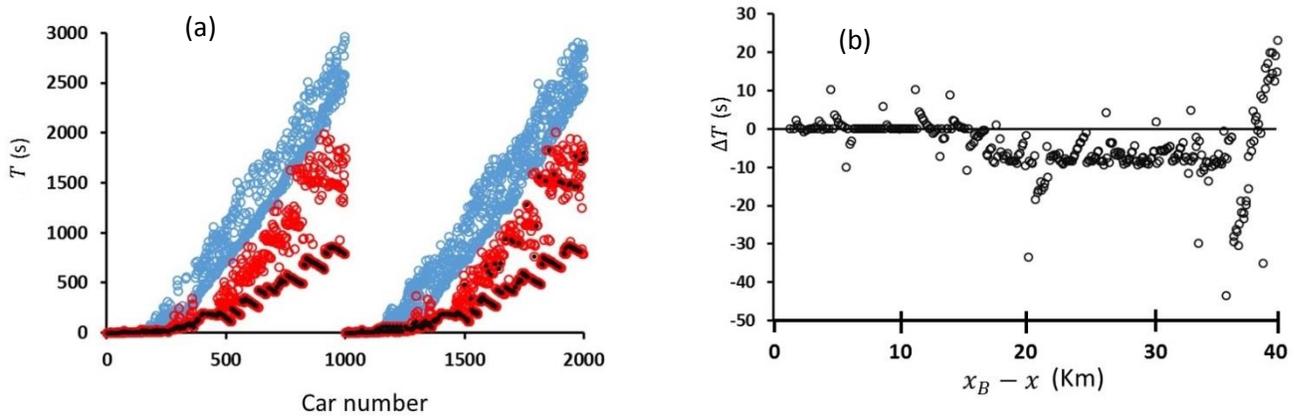

**Fig. 15.** (a) The time delay $T$ for flow of only manual vehicles (blue circles) and flow with 30% wirelessly connected vehicles (red circles, black dots indicate the wirelessly connected vehicles) plotted against car number. Cars 1-1000 (1001-2000) were initially in the continuous (terminated) lane. [A few manual vehicles that appeared to get temporarily stuck near $x_B$ are not shown.] (b) An enlarged view of the difference $\Delta T$ between without and with connected vehicles for cars initially less than 40 Km from $x_B$. Negative values indicate smaller delays for vehicles when there are no wirelessly connected vehicles.

Fig 15b for those initially closer than 40 Km to the bottleneck. Positive $\Delta T$ indicates vehicles that were delayed longer when there was no WCV. In other words, for these vehicles the addition of WCV improved their time to travel past the bottleneck. However, some vehicles (mostly those that are WCV) took slightly longer under system-optimal conditions than if their drivers had optimized their individual times. This is reminiscent of the prisoner's dilemma game structure discussed in [18, 19].

## 6. Conclusions

The simulations reported here demonstrate how wirelessly connected vehicles can be used to reduce congestion at a bottleneck caused by a 2-to-1 lane reduction, even at low penetration (small $f$). As each wirelessly connected vehicle approaches the bottleneck, it receives information from preceding vehicles about slower speeds ahead. Decelerating smoothly causes the following manually driven vehicles to also slow and induces changes to the continuous lane, thereby eliminating or at least prolonging the onset of self-organized congestion. Increased flow out of the congested state is due to the adaptive cruise control feature of the WCV. Other applications of this method can be envisioned, for example, merging at an on-ramp. However, additional control algorithms such as cooperative merging might be required [35, 36].

The congested state, when it occurs, appears quite similar under a variety of conditions. For moderate incoming flow, a transition to the congested state (or not) appears to be related to the sequence of WCV. A large gap between those vehicles that are wirelessly connected (at the same overall percentage penetration) can induce a transition. Likewise, the occurrence of a pre-peak appearing prior to the transition at large incoming flow is also a stochastic event, primarily due to the stochastic properties of the KK model used for the manually driven vehicles.

Calculations for connected vehicles lacking adaptive cruise control show that congestion can be delayed, although the asymptotic flow rate is not increased. A similar effect is observed for WCV with and without deceleration control. Overall, controls are beneficial and give substantially smaller delays for travel past the bottleneck. However, for some vehicles that are initially closer to the bottleneck, the delay times can be slightly larger under the system-optimal conditions.

# Appendix. Kerner-Klenov Model [23-25]

## A.1. *Update rules*

The velocity and position at time step ($t+1$) is calculated for any manual vehicle from its current (at time $t$) velocity $v$ and position $x$ according to the following update rules. The position at $t+1$ is given by

$$x \to x + v_{new} \tag{A.1}$$

and the velocity by

$$v \to v_{new} \tag{A.2}$$

where

$$v_{new} = \min\{v_{limit}, \tilde{v} + \xi, v + 0.5, v_s\}. \tag{A.3}$$

The velocity and position at $t$ of the vehicle immediately in front are denoted by $v_l$ and $x_l$. The gap between the vehicles is

$$g = x_l - x - 7.5. \tag{A.4}$$

The speed limit is $v_{limit}$ and the other quantities are defined below. They are in metric units, *i. e.,* meters, seconds, meter/second, *etc*. Random acceleration and deceleration is given by

$$\begin{aligned}\xi &= 0.5, \quad S = 1 \quad and \quad rnd \le 0.17, \\ &= -0.5, \quad S = -1 \quad and \quad rnd \le 0.1, \\ &= 0, \quad otherwise.\end{aligned} \tag{A.5}$$

The symbol *rnd* represents a random number in [0, 1] and

$$\begin{aligned}S &= -1, \quad \tilde{v} < v - 0.01, \\ &= 1, \quad \tilde{v} > v + 0.01, \\ &= 0, \quad otherwise.\end{aligned} \tag{A.6}$$

where

$$\tilde{v} = \min\{v_s, v_c, v_{limit}\} \tag{A.7}$$

and

$$v_s = \min\{g + v_{la}, v_{safe}\} \tag{A.8}$$

with

$$v_{la} = \max(0, \min\{g_l, v_l - 0.5, v_{lsafe} - 0.5\}) \tag{A.9}$$

where $g_l$ is gap between the lead vehicle and the one in front of it and $v_{lsafe}$ is the "safe" velocity of the lead vehicle analogous to $v_{safe}$ defined below, Eqs. (A.10-16). It is introduced to avoid collisions.

$$v_{safe} = \alpha_{safe} + \beta_{safe} \tag{A.10}$$

and

$$\alpha_{safe} = Int(z) \tag{A.11}$$

where *Int(z)* denotes the integer part of $z$ and

$$\beta_{safe} = \frac{X}{\alpha_{safe} + 1} - 0.5\alpha_{safe}, \tag{A.12}$$

where

$$X = \alpha\beta + 0.5\alpha(\alpha - 1) + g, \tag{A.13}$$

$$\alpha = Int(v_l), \tag{A.14}$$

$$\beta = v_l - \alpha, \tag{A.15}$$

and

$$z = \sqrt{2X + 0.25} - 0.5. \tag{A.16}$$

Furthermore

$$v_c = v + a \quad \text{if} \quad g > G,$$
$$= v + \Delta, \quad otherwise, \tag{A.17}$$

where the synchronization length is

$$G = 3v + 2v(v - v_l) \tag{A.18}$$

and the stochastic time delay of acceleration and deceleration is represented by

$$\Delta = \max\{-b, \min\{v_l - v, a\}\}. \qquad (A.19)$$

To calculate $a$ and $b$, let

$$p_0 = 0.575 + 0.125 \min\{\frac{v}{10}, 1\}, \qquad (A.20)$$

$$p_2 = 0.48, \quad v \leq 15,$$

$$\qquad (A.21)$$

$$= 0.8, \quad v > 15,$$

$$P_0 = 1, \quad S^{prev} = 1,$$

$$\qquad (A.22)$$

$$= p_0, \quad otherwise,$$

and

$$P_1 = p_2, \quad S^{prev} = 1,$$

$$\qquad (A.23)$$

$$= 0.3, \quad otherwise.$$

$S^{prev}$ is the quantity $S$, Eq. (A.6), calculated for this vehicle in the previous time step. Then

$$a = 0.5, \quad rnd < P_0,$$

$$\qquad (A.24)$$

$$= 0, \quad otherwise,$$

and

$$b = 0.5, \quad rnd < P_1,$$

$$\qquad (A.25)$$

$$= 0, \quad otherwise.$$

A.2. *Lane changing rules*

The rules for changing lanes are as follows. The vehicles forming the gap in the target lane are denoted by "ahead" and "follow" while the vehicle immediately in front in the same lane is the "lead" vehicle at $x_l$ with velocity $v_l$. The position and velocity of the vehicle considered for a lane change are $x$ and $v$.

The motivation conditions are:
$$v \geq v_l \tag{A.37}$$
and
$$v_{ahead} \geq v_l + 2. \tag{A.38}$$
Then let
$$g^+ = x_{ahead} - x - 7.5, \tag{A.39}$$
$$g^- = x - x_{follow} - 7.5, \tag{A.40}$$
$$G^+ = \min\{G_{ahead}, v\}, \tag{A.41}$$
and
$$G^- = \min\{G_{follow}, v_{follow}\}, \tag{A.42}$$
where
$$G_{ahead} = 3v + 2v(v - v_{ahead}) \tag{A.43}$$
and
$$G_{follow} = 3v_{follow} + 2v_{follow}(v_{follow} - v). \tag{A.44}$$
The security conditions require
$$g^+ > G^+ \tag{A.45}$$
and
$$g^- > G^-. \tag{A.46}$$
If all conditions are satisfied, a lane change occurs if
$$rnd \leq p_c = 0.45.. \tag{A.47}$$
In the calculations presented in this paper, if the gap ahead in the target lane or the same lane is larger than 150 m, then the speed of the leading vehicle is taken as effectively infinite.